\documentclass{article}
\usepackage{ijcai23}

\usepackage{times}
\usepackage{soul}
\usepackage{url}
\usepackage[hidelinks]{hyperref}
\usepackage[utf8]{inputenc}
\usepackage[small]{caption}
\usepackage{graphicx}
\usepackage{amsmath}
\usepackage{amsthm}
\usepackage{booktabs}
\usepackage{algorithm}
\usepackage{algorithmic}
\usepackage[switch]{lineno}

\title{Learning Cyber Defence Tactics from Scratch with Multi-Agent Reinforcement Learning}

\author{
Jacob Wiebe$^1$\and
Ranwa Al Mallah$^1$\And
Li Li$^2$\and
\affiliations
$^1$Royal Military College of Canada\\
$^2$Defence Research and Development Canada\\
\emails
\{jacob.wiebe, ranwa.al-mallah\}@rmc-cmr.ca,
li.li@ecn.forces.gc.ca
}

\begin{document}

\maketitle

\begin{abstract}
Recent advancements in deep learning techniques have opened new possibilities for designing solutions for autonomous cyber defence. Teams of intelligent agents in computer network defence roles may reveal promising avenues to safeguard cyber and kinetic assets. In a simulated game environment, agents are evaluated on their ability to jointly mitigate attacker activity in host-based defence scenarios. Defender systems are evaluated against heuristic attackers with the goals of compromising network confidentiality, integrity, and availability. Value-based Independent Learning and Centralized Training Decentralized Execution (CTDE) cooperative Multi-Agent Reinforcement Learning (MARL) methods are compared revealing that both approaches outperform a simple multi-agent heuristic defender. This work demonstrates the ability of cooperative MARL to learn effective cyber defence tactics against varied threats.
\end{abstract}
\section{Introduction}

The pace of technological advancements in modern network infrastructure has levied demand for highly skilled and experienced cybersecurity professionals in roles such as penetration testing, threat hunting, and incident response. Meanwhile, machine learning and deep learning have risen as dominant technologies in the space of complex problem-solving. The emerging field of autonomous cyber operations has sought to harness the potential of Reinforcement Learning (RL), deep learning, and multi-agent systems to model the complexity of the cyber battlespace. Autonomous agents operating in cyber defence environments enable rapid tactical-level response to threats given complex observational inputs \cite{vyas2023automated}.

Cyber defence analysts typically employ a chain of decisions leading them from the first discovery of a threat to incident response and mitigation. Human experts are the best tool available for this role. However, human ability in cyber defence tasks faces the challenges of, among others, attention allocation, cognitive load, lack of measurable impact, and reaction time \cite{gutzwiller2015human}. Furthermore, cyber defence is characterized by a large search space of computer network features with which vulnerabilities may be exposed and exploited. To provide coverage over this space, teams of analysts typically coordinate actions. Autonomous RL agents can derive tactical policies in concert to achieve a coordinated effect, mitigating some of the limitations of human actors. This multi-agent approach has yet to be shown in a complex network defence setting.

This work demonstrates the applicability of cooperative Multi-Agent Reinforcement Learning (MARL) for tactical-level decision-making in cyber defence. The results of this work show that cooperative MARL can learn to defend a simulated network environment against three heuristic attack patterns with an emphasis on host-based identification and mitigation of lateral movement.

There are numerous advantages to a cooperative MARL approach over single-agent RL for ACD: (1) input-output spaces can be constrained, avoiding the curse of dimensionality while enabling a high resolution representation of the environment, (2) agents can learn specialized roles relating to a specific function or area of a network, and (3) segregating agents into local areas of control can allow for improved robustness (e.g., if a particular agent is compromised an attacker cannot leak observations from other agents/network segments).

This work makes the following contributions:

\begin{enumerate}
    \item{To the best of our knowledge, CyMARL (Cyber MARL) is the first cooperative MARL training environment for enterprise network defence tasks. It extends the CybORG simulator with additional game types, actions, and network topologies and it provides a PyMARL environment interface with which it is possible to train tens of open-source algorithms.}
    \item{A comparison of cooperative MARL approaches to learn independent and centrally-learned policies in a ACD context.}
    \item{A demonstration of the adaptability of cooperative MARL training when encountering multiple attacker types and action sets.}
\end{enumerate}

The remainder of this paper is organized as follows. Section \ref{cha:related-work} reviews related work in autonomous and multi-agent applications for cyber defence and advancements in cooperative MARL. Section \ref{cha:background} discusses the background of RL, cooperative MARL, and autonomous cyber defence. Section \ref{sec:design} describes the evaluation of the two cooperative MARL approaches tested. Section \ref{sec:discussion} discusses the results. Section \ref{sec:conclusion} suggests future work and concludes.

\section{Related Work}
\label{cha:related-work}

RL for specific cybersecurity tasks has been studied in the context of  DDoS protection \cite{xu2007defending}, anomaly-based intrusion detection \cite{utic2022survey}, and penetration testing \cite{chowdhary2020autonomous,hu2020automated}, among other specific use cases \cite{nguyen2019deep}. RL has been applied to a variety of decision-making tasks relating to cyber defence \cite{wang2022research}. Although ACD techniques have been evaluated using multiple game environments and solving methods \cite{vyas2023automated}, to the best of our knowledge, this work is the first to consider multi-agent coordination of tactical decision-making using state-of-the-art RL techniques.

\subsection{ACD Training Environments} Game environments are commonly modelled as graph-based abstractions of computer networks. Simple defensive decision-making games can be learned using tabular RL methods \cite{hu2020adaptive,applebaum2022bridging}. In more complex game environments, deep RL is preferred for its generalizability and representational capacity. Methods such as DDQN \cite{van2016deep}, A3C \cite{mnih2016asynchronous}, and PPO\cite{schulman2017proximal} have been shown to learn to minimize attacker propagation in graph-based network games modelled as Partially-Observable Markov Decision Processes (POMDP) \cite{han2020adversarial,nyberg2023training}.

More realistic behaviours can be trained using more detailed simulated data. Simulators such as CybORG \cite{standen2021cyborg}, CyberBattleSim \cite{msft:cyberbattlesim}, and FARLAND \cite{molina2021network} emphasize host-based features from which agents learn. Host-based simulations improve task realism by providing more possible states, represented by host processes, vulnerabilities, sessions, and opportunities for Red and Blue agents to interact with these features.
In contrast, Yawning Titan \cite{andrew2022developing} enables greater detail for agents to observe and affect network traffic in simulation.

Foley \textit{et al.} demonstrate how a hierarchical PPO algorithm can learn to adapt to different attacker types to defend a high value server from a heuristic attacker in a simulated environment \shortcite{foley2022autonomous}, winning CAGE Challenge 2 \shortcite{cage_challenge_2_announcement}. Building from this foundation using the CybORG simulator, Wolk \textit{et al.} analyze the generalizability of various RL methods used in CAGE Challenge 2 \cite{wolk2022beyond}.

Emulation has been used to bridge the sim-to-real gap between RL systems that can be trained in defence games and real network operations \cite{li2023enabling,molina2021network}. Emulated environments are impractical for training agents in many research settings due to higher requirements for compute and clock-time than comparable simulations, instead providing a valuable validation component for agents trained in simulation. Red versus Blue asymmetric attack-defence games have used self-play to train defenders against RL-trained attackers \cite{gabirondo2021towards,kunz2023multiagent}. RL self-play and adversarial attacks on RL cyber defenders \cite{standen2023sok} are outside the scope of this research.

\subsection{Multi-Agent Systems for ACD}
\label{sec:mas}

Communication between heuristic defender agents has been shown to improve their ability to protect a simulated network from DDoS attacks \cite{kotenko2007multi}. Heuristic defender agents programmed to communicate filter configurations between teams had improved reaction time and effectiveness than those that did not share information. MARL systems have defended a simulated network against DDoS attacks by selectively throttling network traffic without the use of communication \cite{malialis2015distributed}. MARL systems have successfully performed anomaly-based intrusion detection when framed as a classification task \cite{servin2008multi,shi2021collaborative}.

One of the most commonly used cooperative MARL environments for research is the StarCraft Multi-Agent Challenge (SMAC) \cite{samvelyan2019smac}. SMAC provides a variety of scenarios ranging from easy to very hard in which virtual units learn to defeat each other in a battle video game. A comparative study of cooperative MARL algorithms across a variety of environments, including SMAC, showed that CTDE methods generally perform better than independent learning \cite{papoudakis2020benchmarking}. SMAC tasks are assumed to have transferability for general tactical-level decision-making as a result of agents needing to target actions in a coordinated way toward individual enemies in order to achieve success. The cyber defence problem is modelled similarly, using actions to selectively affect host characteristics.

In practice, the QMIX MARL algorithm \cite{rashid2018qmix} has outperformed state-of-the-art value-based and policy-based methods at a range of tasks \cite{papoudakis2020benchmarking,samvelyan2019smac}. Most notably, QMIX has been shown, with the help of specific implementation tricks, to discover optimal or near-optimal policies on all SMAC \cite{samvelyan2019smac} scenarios \cite{hu2021rethinking}.

\section{Background}
\label{cha:background}

A useful ACD agent must be able to interpret its environment and derive a logical chain of decisions. The RL learner, an autonomous agent, takes sequential actions given inputs from its environment. The agent’s goal is to choose actions that maximize its cumulative reward, referred to as \textit{return}, received from the environment. To learn which series of actions will produce favourable results, an agent must explore the environment, receive rewards, and update its understanding. This process occurs iteratively, allowing the agent to refine its knowledge over many steps and episodes of play. To provide an analysis of how learning architectures compare when used to learn cyber defence tasks, this work constrains the approaches studied to model-free, value-based algorithms.

\subsection{The RL Framework}
\label{sec:rl-framwork}

In a POMDP, an agent will take action $a_t$ causing the environment to step to the next state $s_{t+1}$ based on a transition probability $p$, where $p(s_{t+1} | s_t, a_t)$ is the transition function. The environment generates two outputs from the state transition: the reward $r(s, a)$ as a function of the state and action at time $t-1$, and the observation $o(s)$ as a function of the state at time $t$. The observation is some subset of state information determined by the observation function. A \textit{game} in this context generalizes the POMDP and refers to an environment with a consistent set of rules in which one or more agents interact.

An agent’s \textit{policy} $\pi(a|o)$ is a mapping of actions to the observation space. The policy, therefore, determines a series of actions for an agent to take. An agent adjusts its policy over many episodes of a game to optimize for its cumulative reward in an episode, its \textit{return} \cite{sutton2018reinforcement}.

A value-based agent will predict the value of states or state-action pairs using its value function and decide on an action that will maximize its expected future return $\pi(o) = \arg\max_{a \in A} Q(o, a)$. The value function is calculated from the expected discounted future return:

\begin{equation}
	Q(o, a) =  \alpha[r + \gamma \max_{a_{t+1}} Q(o_{t+1}, a_{t+1})] + (1 - \alpha)Q(o, a)
	\label{eq:q-update}
\end{equation}

Where $\alpha$ is the learning rate and the discount rate $(\gamma \in (0,1))$ weights rewards less as the agent looks to further $k$ states. To encourage exploration, rather than to continue to revisit known high-value states, an $\epsilon$-greedy strategy is commonly employed which sets a probability $\epsilon$ that an agent will randomly select an action rather than taking the action with the highest expected reward.

With deep value-based methods, sampling from experience replay \cite{lin1992self} often produces more favourable results, as demonstrated by Mnih \textit{et. Al} with Deep Q-Networks \shortcite{mnih2013playing}. A replay buffer can improve sample efficiency by exposing the agent to more learning experience without generating each sample through actions. Recurrent Neural Networks (RNN) allow RL models to learn long-term dependencies between data samples which is useful for accurate predictions in partially observable environments \cite{bakker2001reinforcement}.

\subsection{Multi-Agent Reinforcement Learning}
\label{sec:marl}

\begin{figure}
    \centering
    \includegraphics[width=0.95\linewidth]{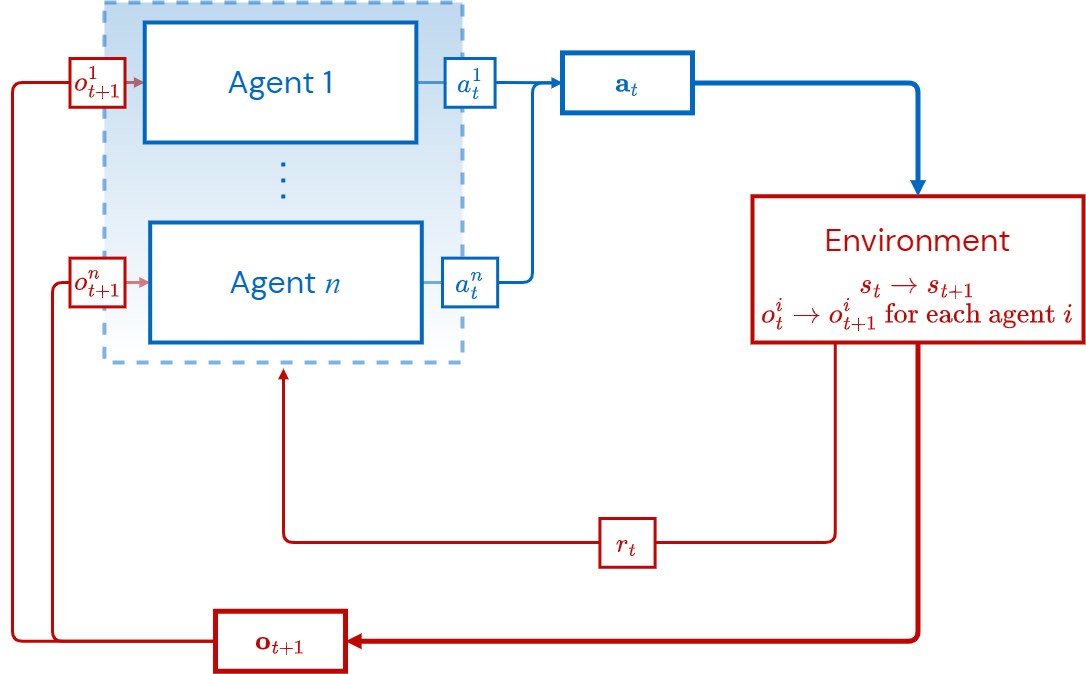}
    \caption{Multi-Agent RL Agent-Environment Framework.}
    \label{fig:mdp-marl}
\end{figure}

The cooperative multi-agent game builds upon the POMDP defined previously by allowing multiple agents to take actions and receive rewards simultaneously within a single timestep \cite{littman1994markov,bucsoniu2010multi}. Although each agent, denoted by $i$, receives individual observations $o^i$ from the environment, a joint reward $r$ is determined based on the state and actions of all agents. Agents collectively seek to optimize the joint reward. Figure \ref{fig:mdp-marl} provides a cooperative MARL representation of agent-environment interaction. The game’s global state depends on the vector of actions $\textbf{a}_t=\{a^{1}_{t},...,a^{n}_{t}\}$, where $i=1,2,...n$ for $n$ agents. At each timestep, the environment will output a joint reward as a scalar and an observation vector $\textbf{o}_{t+1} = \{o^{1}_{t+1}, ...,o^{n}_{t+1}\}$ that includes individual observations for each agent.

Simultaneous actions in a cooperative MARL setting allow for greater exploration. Cooperative MARL also has the advantage of dividing large tasks into manageable goals, thus handling greater task complexity. However, since the game’s current state depends on the joint actions of all agents, an agent’s actions can affect other agents’ observations. Since both agents are learning and updating their policies, this interaction creates \textit{multi-agent}-non-stationarity in the environment; the optimal policy is a moving target. In a single-agent game, the fixed probability of a stochastic transition function can be incorporated into the learned policy, for example, with the use of a replay buffer. However, in the multi-agent case, actions of “external agents” affect observations in a way that cannot be explained by changes in the observing agent’s policy \cite{lowe2017multi}. It can therefore be challenging for an agent to learn a stable policy when the actions of its peers are silently affecting its reward and observations.

\subsubsection{Independent Q-Learning}

A naive approach to cooperative multi-agent learning systems is to decentralize prediction and control, referred to as \textit{independent learning}. This approach employs self-contained RL agents that act out separate policies based on independent observations while seeking to maximize a joint reward. Each agent is responsible for learning a policy from their individual observations. A decentralized approach allows for greater scalability, at the cost of high variance due to the unmitigated non-stationarity. Independent Q-Learning (IQL), as implemented in this work, is equivalent to training independent DQN-style agents\footnote{This work utilizes an implementation of IQL from PyMARL2 \cite{hu2021rethinking} in which each DQN-style agent uses an RNN in addition to other implementation tricks to optimize for cooperative play.} to learn to take simultaneous actions to maximize their joint reward.

\subsubsection{QMIX}
\label{ssec:qmix}

QMIX \cite{rashid2018qmix} is a Centralized Training with Decentralized Execution (CTDE) \cite{lowe2017multi} algorithm that uses a central mixing architecture to perform \textit{value decomposition} \cite{sunehag2017value}. With CTDE, semi-independent agents follow separate policies and receive updates periodically from a central learner. The central learner trains on information from all agents and provides a learning update, allowing each agent’s policy to condition on the policies of their peers, mitigating the non-stationarity problem. The foundational difference between QMIX and IQL implementations in this work is that QMIX performs learning updates centrally. 

\subsection{Autonomous Cyber Defence}
\label{sec:acd}
ACD does not seek to replace tools that are successful in the field, such as anomaly-based intrusion detection. It instead aims to build a tactical-level decision-making framework to integrate with existing technologies. In this research, the decision space is modelled as a simulation on top of a tool-based abstraction of the network state. An RL simulation for ACD must model the elements of the environment to the level of detail that will allow the agent to learn tactical-level action chains when presented with a series of alerts about the underlying network status. A multi-agent framework further allows for the decentralization of the decision-making processes while maintaining an input-output space that is constrained to a level that can be learned effectively by current RL methods.

The high-level actions of the defender in this game represent the choices for a defender to selectively understand or act against the threat. The tactical decision-making competency of agents then can be inferred from their ability to minimize the impact of the attack. Many thousands or millions of iterations are often required for RL to learn reasonable policies. This problem is exasperated by a lower sample efficiency of partially-observable environments \cite{papadimitriou1987complexity}. By simulating ACD, RL systems can be trained over many iterations in an abstraction of the underlying computer network environment. Simulation requires less compute than the alternative emulated approach while still offering enough complexity to challenge current state-of-the-art designs.

Cyber Operations Research Gym (CybORG) is a framework that provides a simulated cyber operations environment for training and evaluating RL agents \cite{standen2021cyborg}. CybORG simulates information technology systems related to network security that a cybersecurity professional may use in network defence or penetration testing. The simulated environment presented in this work, CyMARL, adapts CybORG to create a multi-agent host-based monitoring game for training tactical cyber defence. CyMARL includes a PyMARL\footnote{PyMARL is an open-source MARL research project that allows algorithms to be built from existing deep RL components such as agents and trainers. It is included as part of the SMAC environment and paper \cite{samvelyan2019smac}. The repository for PyMARL can be found at https://github.com/oxwhirl/pymarl.} environment allowing for integration with the tens of open-source cooperative MARL algorithms built on the PyMARL framework.

\section{Experimental Design}
\label{sec:design}

To demonstrate the applicability of cooperative MARL for tactical-level decision-making for ACD, three game scenarios are evaluated against three attacker types. CAGE Challenge 2 \cite{cage_challenge_2_announcement} is a central contributor to many aspects of this experiment, including the attacker models, network topology, and agent actions. CyMARL extends the complexity of CAGE Challenge 2 by employing new scenarios (with diverse objectives, attacker behaviours, and reward functions), modified observation and action spaces, and a cooperative multi-agent framework.

\subsection{Simulated Network Topology}

The simulated game environment is composed of nine hosts, five in a user subnet and four in an operational subnet. Hosts are a mix of Linux and Windows, with unique exploits for each OS. Hosts simulate processes, users, active sessions, and network interfaces. Rewards are calculated from the importance score of hosts. Hosts in the user subnet have a score of 0.1, operational subnet hosts, 1, and the operational server (located in the operational subnet), 10. Each host in a \textit{compromised} state incurs a negative reward for the defender team equal to its importance score.

\begin{figure*}
	\centering
	\includegraphics[width=0.55\linewidth]{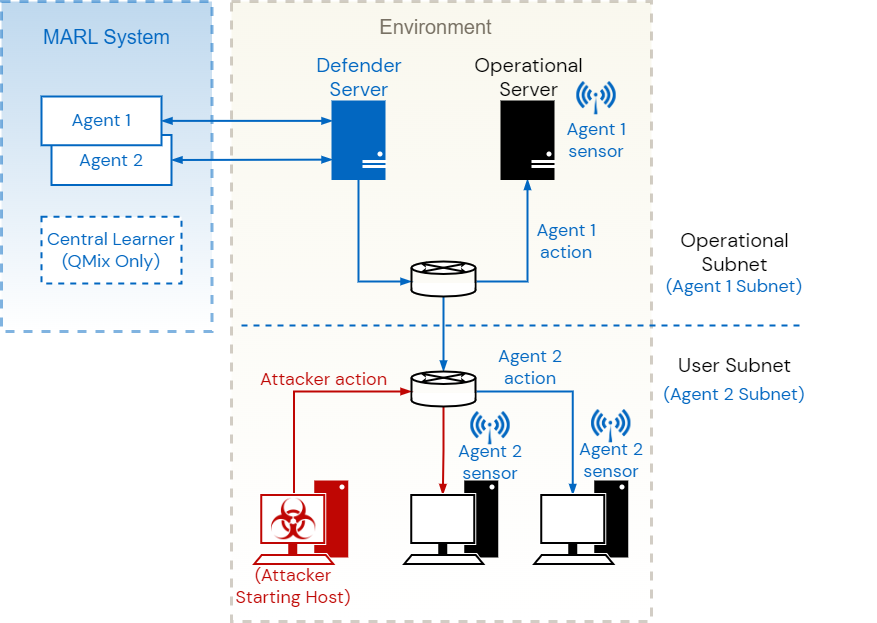}
	\caption{Simple network diagram with defender sensors and targets.}
	\label{fig:blue-net}
\end{figure*}

A heuristic attacker agent begins each episode with an initial foothold on a user host and moves between hosts by scanning for and exploiting vulnerable services. The attacker can move laterally to a target host if one of its sessions' hosts has visibility of the target host through its network interface. The attacker's actions follow a hierarchy: it must first discover a host, then a port, before attempting to move laterally on the network using an exploit. Escalating session privilege can then be performed. The attacker’s session cannot be removed from the initially compromised host allowing for episodic gameplay where both sides have opportunities to influence the reward. RL defender agents are segregated by subnet, one agent responsible for the user subnet and the other controlling the operational subnet. An example of the agent-network interface is shown in Figure \ref{fig:blue-net}. 

\subsection{Attack Patterns}

A real-world attacker is often not solely motivated to establish communication channels on a network. The heuristic attacker's behaviour in this game is differentiated by its goal, which can be \textit{confidentiality}, \textit{integrity}, or \textit{availability}. In each case, the attacker takes different actions upon establishing a session on a host, though the heuristic for lateral movement is the same across all scenarios. In the confidentiality scenario, the attacker's intent is to spy on privileged information within the network but not take actions to modify it directly. The confidentiality attacker achieves “compromise” for each host it is connected to at each timestep. The integrity attacker seeks to modify files stored on a host. At each timestep that modified data exists, a host is considered compromised. The availability attacker will spawn a Denial of Service (DoS) malware process that represents a consumption of CPU cycles, affecting user availability. Similarly, each host in this scenario with a malware process running is compromised. This does not affect the defender's ability to monitor the victim host. Although the attacker's behaviour is a simplification of real attack patterns, the difference in objective affects how the threat appears to the defender and which actions it should take.

To compromise a host in the integrity and availability scenarios, an additional action must be taken once a session is established on a victim host using the \texttt{exploit} action. The \texttt{tamper} action is taken by the integrity attacker to simulate the modification of important files. The \texttt{deny} action is taken by the availability attacker to simulate the creation of a DoS process. The two-step scoring process in the integrity and availability scenarios decouples the observation of attacker presence on a host from the signal of a host compromise via the negative reward.

The cumulative reward for each strategy serves to model the risk to the network while malicious artifacts exist on network hosts. As the number of timesteps increases, there is a greater probability that the attacker will have collected privileged information or disrupted normal processes on the network. Likewise, the severity of the risk is proportional to the importance of the host that becomes compromised.

\subsection{Observations}

Host-based monitoring systems are a common and effecti tool for cyber defence operations, particularly in situations where an attacker has made an initial breach and is attempting to gain Command and Control (C2) channels within a network. In CyMARL, each defender agent’s observations are modelled as an abstraction of information that may be provided from a host-based monitoring service. The defender receives a set of six flags at each turn for each host (denoted by the variable $x$) within its area of control:

\begin{enumerate}
	\item Host $x$ was scanned last turn.
	\item An exploit was run on a Host $x$ service last turn.
	\item There are no malicious sessions on Host $x$.
	\item Host $x$ status is unknown.
	\item A malicious session on Host $x$ exists with user privileges.
	\item A malicious session on Host $x$ exists with root/admin privileges.
\end{enumerate}

The flags are one-hot encoded into a vector that is assigned to an agent ID for interface with PyMARL. These observations do not specifically alert of malicious files or processes so the defender must learn patterns between the information encoded within its observation space and its rewards.

\subsection{Defender Actions}
To mitigate attacker activity, the defender agents select actions to monitor, remove suspicious processes or files, restore hosts, or gather file information. \texttt{monitor} returns the observation vector and checks all hosts within an agent's vision for new processes that are then added to a suspicious process list. An agent can then choose to \texttt{remove} suspected malicious processes from a specific host. As a more aggressive action, an agent can re-image a host using \texttt{restore}. Restoring is the only way to remove an attacker that has escalated its privilege on a host. It will also remove any malicious processes or files. However, the \texttt{restore} action has an associated penalty that scales with the importance score of the target host.

In the integrity scenario, agents have additional actions to manage the specific threat type. Agents can take an \texttt{analyze} action to determine the security characteristics of files on a host, allowing it to identify any malicious files. To remove the discovered files, the \texttt{data\_repair}, is required. In the availability scenario, the \texttt{remove} action can be used to target DoS processes and simulated TCP sessions. In the second set of trials, the \texttt{misinform} action is introduced which generates a service on a host that can be misinterpreted by an attacker as vulnerable. The attacker may choose to exploit this process which will always fail. A defender agent may choose to perform any action on any of its connected hosts. However, some actions are invalid (e.g., if a host is targetted with the remove action but does not have a malicious process running) and are penalized with a reward of -0.1.

\subsection{Heuristic Defender}

To validate the performance of MARL systems, a heuristic multi-agent defender is used as a benchmark for each game scenario. The heuristic defender's design assumes it has access to a tailored expert system that provides a precise signal of malicious process creation. It relies on selected information from the true state and does not utilize the observation space of the MARL systems. In the confidentiality scenarios, this policy monitors all hosts within each agent’s subnet, and restores hosts indicated by the process creation alert. The \texttt{restore} action was chosen because it scored higher in testing than using combinations of other actions.  In the availability scenario, detection occurs due to a DoS process being created, not an attacker session. In the integrity scenarios, the heuristic defender will attempt to remove malware on a host that it has discovered to have a tampered file with a 50\% probability and restore otherwise. In trials using the \texttt{misinform} action, the defender system creates a decoy process on a random host for the first five turns and then proceeds with its standard policy for the remainder of the episode. Since using \texttt{monitor} to detect and \texttt{restore} to re-image affected hosts takes two turns, the attacker has the opportunity to stay one step ahead if it continues to exploit new hosts. Although the heuristic defender has perfect game information, its policy is restrictive enough to allow the attacker opportunities to overcome its defences.

\subsection{Evaluation Metrics}

For each experiment, two algorithms are evaluated: IQL and QMIX, implementing fully-decentralized and CTDE architectures, respectively. Agents do not directly share information with each other, but in the case of QMIX, a joint policy representation is learned centrally before being decomposed into separate behaviour policies. The architecture of cooperative MARL systems has been shown to significantly affect the learning ability of models at a range of tasks \cite{papoudakis2020benchmarking,rashid2018qmix}. Learning ability is the capacity of an RL model to generate a policy from its initial randomized state. A trained model that achieves a higher evaluation score (i.e., mean return) at a task has a higher learning ability than a lower-performing model given the same training opportunity.

MARL systems are trained for two million timesteps. A timestep is represented in the game as a turn in which each agent takes one action. Each trained model is evaluated over 1000 timesteps of play without learning. The evaluation score is the average of the mean return of five evaluation runs using separately trained static policies for each trial. Evaluation scores are compared using percent difference. The variance of models is expressed in the standard deviation of the mean score. Learning speed is an implicit evaluation criterion to mitigate the impracticality of long training times. Although there may be greater performance gains outside of the bounds of the training time for the model, training time is constrained to conserve computational resources. This allows for greater ease of reproducibility of these results and a reasonable expectation of performance given a fixed amount of model experience.

\begin{figure*}[t]
	\centering
	\includegraphics[width=0.95\linewidth]{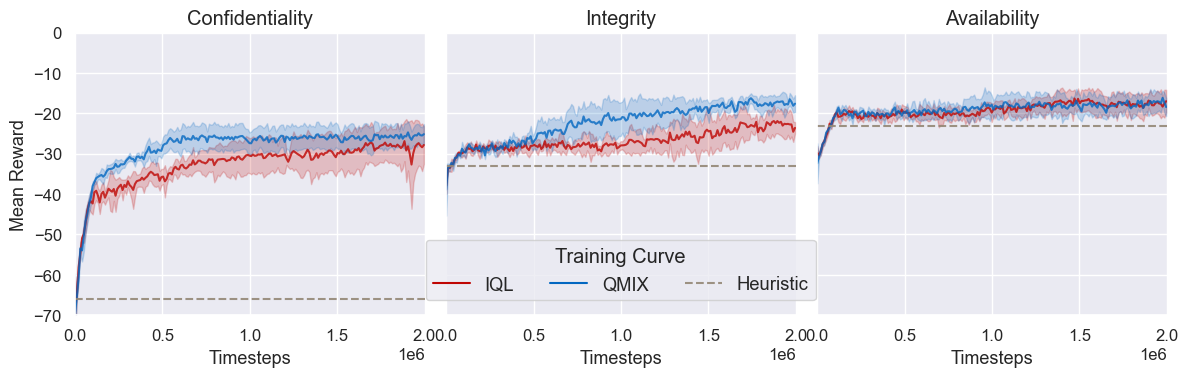}
	\caption{Learning curves of IQL and QMIX compared to heuristic at three scenarios.}
	\label{fig:baseline_trg}
\end{figure*}

\begin{figure*}
	\centering
	\includegraphics[width=0.95\linewidth]{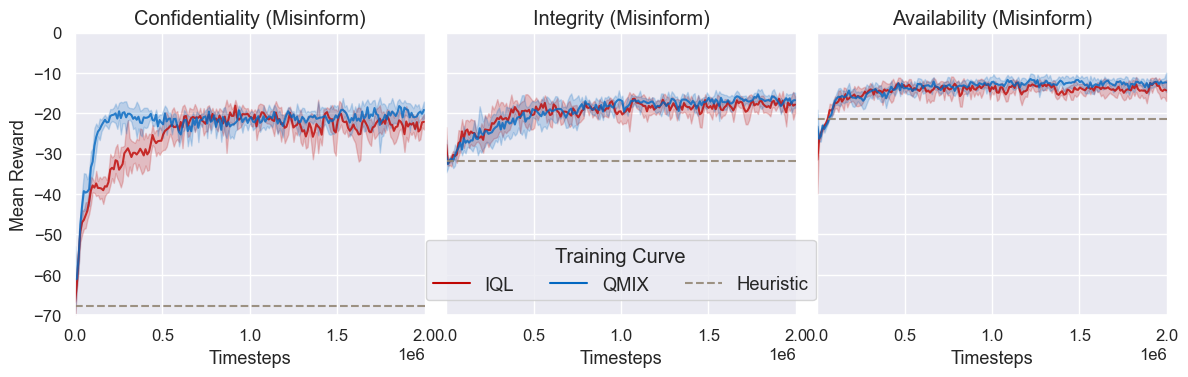}
	\caption{Learning curves of IQL and QMIX compared to heuristic with the addition of the proactive misinform action.}
	\label{fig:misinform_trg}
\end{figure*}

\begin{table*}[t]
  \caption{Mean return +/- standard deviation of trained iql and qmix policies relative to the heuristic model.}
  \label{tab:results}
  \centering
  \begin{tabular}{|l|c|c|c|}
	  \hline
    	\multicolumn{1}{|c}{Scenario} & IQL & QMIX & Heuristic \\
        \hline
    	Confidentiality & -26.19 ±5.5 & \textbf{-23.01 ±2.47} & -65.96 ±24.33 \\
        \hline
    	Integrity & -24.38 ±5.44 & \textbf{-16.8 ±2.73} & -32.97 ±20.82 \\
        \hline
    	Availability & \textbf{-15.94 ±3.78} & -17.5 ±4.68 & -23.11 ±18.88 \\
        \hline
    	Confidentiality (Misinform) & \textbf{-19.76 ±4.18} & -25.07 ±7.1 & -67.64 ±24.03 \\
        \hline
    	Integrity (Misinform) & -18.17 ±3.86 &\textbf{-16.75 ±3.48} & -31.9 ±20.73 \\
        \hline
    	Availability (Misinform) & -14.47 ±2.98 & \textbf{-11.63 ±2.84} & -21.39 ±16.95 \\
        \hline
        \multicolumn{4}{l}{*The highest scores for each scenario are given in bold.}
  
\end{tabular}
\end{table*}

\section{Experiments}

Two architectures are trained and compared at controlling two defender agents in three scenario types each with different attacker behaviour (confidentiality, integrity, and availability) in two sets of experiments. The first set uses the standard action set described in Section \ref{sec:design} and the second adds the \texttt{misinform} action to each scenario. The addition of a proactive action to the set of reactive actions sets the misinform experiment apart and allows for an evaluation of how the two MARL architectures perform under varied conditions.

To set initial conditions, a selection of hyperparameters was varied in a grid search in which MARL systems were trained to 500,000 timesteps at the confidentiality scenario evaluating four parameters: batch size (128, 256), buffer size (5,000, 10,000), learning rate (0.005, 0.01), and TD-$\lambda$ trace decay (None, 0.6). For each set of hyperparameter values, three random seeds were trained to provide a representative result. The simplest scenario is confidentiality scenario due to the attacker’s exploit and the reward for compromise occurring simultaneously. It is assumed that these hyperparameter values transfer reasonably well to the other scenarios. The hyperparameter values chosen were based on initial trials with the environment and the findings in Hu \textit{et al.} and Hessel \textit{et al.} that evaluated the effect of RL implementation designs in QMIX and DQN, respectively \cite{hu2021rethinking,hessel2018rainbow}.

The training curves of IQL and QMIX are shown in Figure \ref{fig:baseline_trg} for the three baseline scenarios and Figure \ref{fig:misinform_trg} for the misinform scenarios. The shaded area of each curve is the standard deviation of the average training return. The dashed line is the mean heuristic defender score for each scenario. Both IQL and QMIX outperform the heuristic multi-agent defender model in all scenarios. QMIX has a slight learning speed advantage in the confidentiality and integrity scenarios but otherwise, both architectures tend towards similar return scores. Table \ref{tab:results} presents the evaluation scores of IQL, QMIX, and the multi-agent heuristic defender at each scenario.

The option to generate decoy processes generally improved the performance of both architectures with a 19.8\% and 8.3\% average improvement in score for IQL and QMIX, respectively. However, QMIX had reduced performance in the confidentiality scenario and a negligible difference in the integrity scenario. The heuristic model using the \texttt{misinform} action at the start of each episode performed 2.7\% better on average.

\section{Discussion}
\label{sec:discussion}

The availability and integrity scenario types employ an attacker that must perform an additional action upon establishing a session to score points. The attacker in these scenario types is therefore slower and will tend to score less in a fixed-timestep game. On average the MARL systems trained at the integrity and availability scenarios respectively scored 19.1\% and 36.7\% higher than the confidentiality scenario. This effect is more pronounced in the heuristic, scoring 51.4\% and 66.7\% higher in the integrity and availability scenarios, respectively, than in the confidentiality scenario. Moreover, the addition of the \texttt{misinform} action narrowed the performance gap between IQL and QMIX from a 11.2\% QMIX advantage to 0.2\%. These results suggest that there is an attractive local minimum for each scenario that both architectures tend to settle in.

The multi-agent host-monitoring game presented does not require explicit coordination for a defender team to be successful. Moreover, the segregation of agents into subnets limits their ability to influence the state of other agents. As a result, IQL is able to discover policies that score closely to those of QMIX despite it's disadvantages of lacking coordination and suffering from multi-agent non-stationarity. IQL also has the advantage of scalability. QMIX requires that Q-functions are learned centrally, thus limiting the number of possible agents since the central learner will eventually create a bottleneck. IQL does not require any central learning and therefore the number of agents in a game does not pose a constraint.

\section{Conclusion and Future Work}
\label{sec:conclusion}

Tactical decision-making in cyber defence contends with an expansive problem space necessitating expert knowledge and, in many cases, decentralization of effort to provide essential coverage. This work presented the initial evidence of the success of cooperative MARL to generate of decentralized, tactical-level control policies for a variety of host-based ACD  scenarios. Independent and CTDE value-based cooperative MARL architectures can learn policies that outperform a basic heuristic model in this game. In particular, the performance of IQL at these tasks suggests that this approach may scale well to larger networks with more agents. Future gains in performance are expected to be possible both in terms of the learning ability of MARL systems and in the realism of training games. We suggest the following areas for the future development of cooperative MARL in the domain of ACD:
\begin{enumerate}
	\item{Techniques to approximate greater environmental realism, including the use of generative programs to increase the sample size of experimentation \cite{ellis2022smacv2,molina2021network},}
	\item{Games offering more sophisticated and varied threat types using self-play \cite{hammar2020finding} and adversarial machine learning attacks \cite{standen2023sok,molina2021network,han2020adversarial}, and}
	\item{Leveraging promising MARL techniques such as multi-agent policy-gradient methods \cite{yu2021surprising} or hierarchical role assignment \cite{wang2020rode}, transfer learning \cite{wang2020few}, attention mechanisms \cite{yang2020qatten} and transformers for value-decomposition \cite{khan2022transformer}}.
\end{enumerate}

\bibliographystyle{named}
\bibliography{ijcai23}

\end{document}